\documentclass{article} 

\newbox\tempa
\newbox\tempb
\newdimen\tempc
\def\mud#1{\hfil $\displaystyle{\mathstrut #1}$\hfil}
\def\rig#1{\hfil $\displaystyle{#1}$}
\def\irulehelp#1#2#3{\setbox\tempa=\hbox{$\displaystyle{\mathstrut #2}$}%
                        \setbox\tempb=\vbox{\halign{##\cr
        \mud{#1}\cr
        \noalign{\vskip\the\lineskip}
        \noalign{\hrule height 0pt}
        \rig{\vbox to 0pt{\vss\hbox to 0pt{${\; #3}$\hss}\vss}}\cr
        \noalign{\hrule}
        \noalign{\vskip\the\lineskip}
        \mud{\copy\tempa}\cr}}
                      \tempc=\wd\tempb
                      \advance\tempc by \wd\tempa
                      \divide\tempc by 2 }
\def\irule#1#2#3{{\irulehelp{#1}{#2}{#3}
                     \hbox to \wd\tempa{\hss \box\tempb \hss}}}
\def\birulehelp#1#2#3{\setbox\tempa=\hbox{$\displaystyle{\mathstrut #2}$}%
                        \setbox\tempb=\vbox{\halign{##\cr
        \mud{#1}\cr
        \noalign{\vskip\the\lineskip}
        \noalign{\hrule height 0pt}
        \phantom{$#3$}
                \rig{\vbox to 0pt{\vss\hbox to 0pt{${\; #3}$\hss}\vss}}\cr
        \noalign{\hrule}
        \noalign{\vskip\the\lineskip}
        \mud{\copy\tempa}\cr}}
                      \tempc=\wd\tempb
                      \advance\tempc by \wd\tempa
                      \divide\tempc by 2 }
\def\birule#1#2#3{{\birulehelp{#1}{#2}{#3}
                     \hbox to \wd\tempa{\hss \box\tempb \hss}}\phantom{#3}}

\usepackage{stmaryrd,amsfonts,amssymb}

\begin{document}
  \title{On the convergence of reduction-based and model-based methods in proof theory} 
\author{Gilles Dowek\thanks{
\'Ecole polytechnique and INRIA, 
    LIX, \'Ecole polytechnique, 91128 Palaiseau Cedex, France,
{\tt gilles.dowek@polytechnique.edu}.}}
\date{}
\maketitle
\thispagestyle{empty}

\begin{abstract}
In the recent past, the reduction-based and the model-based
methods to prove cut elimination have converged, so that they now
appear just as two sides of the same coin. This paper details some of
the steps of this transformation.
\end{abstract}


\section*{Introduction}

Many results of proof theory, such as unprovability results,
completeness of various proof-search methods, the disjunction and the
witness property of constructive proofs and the possibility to extract
programs from such proofs, rely on cut elimination theorems, asserting
that when a proposition is provable in some theory, it has a proof of
a special form: a {\em cut free} proof.

The methods developed to prove such cut elimination theorems can be
broadly divided into two categories: the reduction-based methods and
the model-based ones. In the recent past, we have witnessed a
convergence of these two kinds of methods.  In this paper, I detail
some of the steps of this transformation that has lead to consider
reduction-based and model-based methods just as two sides of the same
coin.

\section{The problem of the axioms}

A preliminary step to the convergence of reduction-based and
model-based methods for proving cut elimination has been the
definition of a sufficiently general notion of cut.  And this
has itself required a modification of the notion of theory.

Let us start with an example.  A consequence of the cut elimination
theorem is the disjunction property: in constructive natural
deduction, when a proposition of the form $A \vee B$ is provable in
the empty theory, {\em i.e.} without axioms, then either $A$ or $B$ is
provable.  Indeed, by the cut elimination theorem, if the proposition
$A \vee B$ is provable, it has a cut free proof. Then, a simple
induction on proof structure shows that, in constructive natural
deduction, the last rule of a cut free proof in the empty theory is an
introduction rule.  Thus, a cut free proof of $A \vee B$ ends with an
introduction rule of the disjunction and either $A$ or $B$ is
provable.

The fact that the last rule of a cut free proof is an introduction
rule does not extend when we add axioms. For instance, if the
proposition $A \vee B$ is an axiom, then it has a cut free proof that
ends with the axiom rule.  Thus, the disjunction property for theories
such as arithmetic, simple type theory, or more simply for the theory
formed with the axiom $P \Leftrightarrow (Q \Rightarrow R)$, cannot be
derived, in a simple way, from the cut elimination theorem for
predicate logic.  The attempts to characterize axioms, such as Harrop
formulae, that preserve the usual properties of cut free proofs, such
as the disjunction property, the witness property or the fact that the
last rule is an introduction rule, have led to relatively small
classes, that, for instance, never contain the axioms of arithmetic.
Indeed, the cut elimination theorem for predicate logic without axioms
can be proved in arithmetic and the property that all propositions
provable in arithmetic have a proof ending with an introduction rule
implies the consistency of arithmetic. Thus, this result cannot be
derived in an elementary way from the cut elimination theorem of
predicate logic.

Therefore, to prove the disjunction property for arithmetic or simple type
theory, we need to extend the cut elimination theorem first. This
explains why there are several cut elimination theorems for various
theories of interest. For instance, in arithmetic, we usually
introduce a new form of cut, specific to the induction axiom, and we
prove the cut elimination theorem again for this extended notion of
cut.

This necessity to introduce a specific notion of cut for each
theory of interest has been an obstacle to the development of a general
theory of cut elimination. {\em Deduction modulo} \cite{DHK,DowekWerner} has
been an attempt to partially solve this problem.  In deduction modulo,
the axioms of the form $\forall x_1~...~\forall x_n~(t = u)$ or $\forall
x_1~...~\forall x_n~(P \Leftrightarrow A)$ where $P$ is an atomic
proposition are replaced by rewrite rules $t \longrightarrow u$ or $P
\longrightarrow A$. For instance, the axiom 
$P \Leftrightarrow (Q \Rightarrow R)$ is replaced by the rewrite rule 
$P \longrightarrow (Q \Rightarrow R)$.
Then, deduction is performed modulo the congruence
generated by these rules. For instance, with the rewrite rule 
$P \longrightarrow 
(Q \Rightarrow R)$, an instance of the elimination rule of the implication 
is 
$$\irule{\Gamma \vdash P~~~\Gamma \vdash Q}
        {\Gamma \vdash R}
        {\mbox{$\Rightarrow$-elim}}$$
because $P$ is congruent to $Q \Rightarrow R$. 
If not all, many theories can be formulated
as rewrite systems: for instance arithmetic
\cite{Peano,Allali} and simple type theory \cite{DHKHOL}.

A cut, in deduction modulo, is defined as in natural deduction: it is a
sequence formed with an introduction rule followed by an elimination
rule. In deduction modulo, not all theories have the cut elimination
property. For instance, the theory formed with the rule $P
\longrightarrow (Q \Rightarrow R)$ does,
as well as the theory formed with the rule $P \longrightarrow (Q \Rightarrow
P)$, but not that formed with the rule $P \longrightarrow (P
\Rightarrow R)$. In deduction modulo, a cut free proof in a purely
computational theory, {\em i.e.} a theory containing rewrite rules but no
axioms, always end with an introduction rule. Thus, when a purely 
computational theory has the cut elimination property, it has the 
disjunction property. 

As a cut in deduction modulo is just a sequence formed with an
introduction rule and an elimination rule, the definition of the
notion of cut is independent of the theory of interest and asking if
some theory, formulated as a rewrite system, has the cut elimination
property is now a well-formed question.

In fact, a similar idea had been investigated earlier by Dag Prawitz
and others \cite{PrawitzND,Crabbe74,Crabbe91,Hallnas,Ekman} who have
proposed to replace the axioms of the form $\forall x_1~...~\forall
x_n~(P \Leftrightarrow A)$ by non logical deduction rules allowing 
to fold $A$ into $P$ and to 
unfold $P$ into $A$.
For instance, the axiom $P \Leftrightarrow (Q \Rightarrow R)$ can be replaced 
by two rules 
$$\irule{\Gamma \vdash Q \Rightarrow R}
        {\Gamma \vdash P}
        {\mbox{fold}}$$
and
$$\irule{\Gamma \vdash P}
        {\Gamma \vdash Q \Rightarrow R}
        {\mbox{unfold}}$$ 
More recently, Benjamin Wack has introduced a notion of super-natural
deduction \cite{Wack,KirchnerBraunerHoutmann}, where the folding of
$A$ into $P$ preceded by all the possible introduction rules of the
connectors and quantifiers of $A$ and the unfolding of $P$ into $A$ is
automatically followed by the corresponding elimination rules.
For instance, the axiom $r:(P \Leftrightarrow (Q \Rightarrow R))$ 
is replaced by the rules 
$$\irule{\Gamma, Q \vdash R}
        {\Gamma \vdash P}
        {\mbox{$r$-intro}}$$
and
$$\irule{\Gamma \vdash P~~~\Gamma \vdash Q}
        {\Gamma \vdash R}
        {\mbox{$r$-elim}}$$
This leads to very natural deduction rules where, like in the usual
mathematical practice, connectors and quantifiers almost disappear.

Interestingly, the theories that have the cut elimination property in
these three formalisms are the same
\cite{foldunfold,BraunerDowekWack}. This shows the robustness of this
notion of cut.

To conclude, it is possible to formulate a general notion of cut that
applies to all the theories that can be expressed as a rewrite system
and this notion of cut is quite robust.  The definition of such a
notion of cut independent of the theory of interest has been a
pre-requisite to the convergence of reduction-based and model-based
methods as, this way, a sufficiently large class of problems to which
these methods apply, has been identified.

\section{Models} 

\subsection{Soundness and completeness}

The soundness theorem asserts that, for every proposition $A$, if $A$
is provable in predicate logic, then for every model ${\cal M}$, $A$
is valid in ${\cal M}$.  Conversely, the completeness theorem asserts
that, for every proposition $A$, if for every model ${\cal M}$, $A$ is
valid in ${\cal M}$, then $A$ is provable.  In most of the proofs of
the completeness theorem however, the classically equivalent statement
is proved: for every proposition $A$, there exists a model ${\cal M}$
such that if $A$ is valid in ${\cal M}$, then $A$ is provable.

With the usual notion of bi-valued model, it is not
possible to permute the quantifiers and to prove that there
exists a uniform model ${\cal M}$ such that for every proposition $A$,
if $A$ is valid in ${\cal M}$, then $A$ is provable.  
But, if we extend the notion
of model by allowing truth values to be elements of an arbitrary
boolean algebra, then this statement becomes provable. An example of
such a model is the model where the term interpretation domain is the 
set of terms of the language and the propositions interpretation domain is 
the Lindenbaum algebra of the language, {\em i.e.} the set of 
proposition of the language quotiented by the relation $\simeq$ defined by 
$A \simeq B$ if $A \Leftrightarrow B$ is provable. 
In constructive
logic, the boolean algebras need to be replaced by Heyting algebras,
but this uniform completeness theorem can still be proved.

\subsection{Model-based cut elimination proofs}

The model-based cut elimination proofs ---~for instance 
\cite{Tait66,Prawitz,Takahashi,Andrews,DeMarcoLipton,Okada}~--- proceed by
proving a sharpened 
completeness theorem: for every proposition $A$, if for every model
${\cal M}$, $A$ is valid in ${\cal M}$, then $A$ has a cut free proof.
The cut elimination theorem is then just a consequence of the
soundness theorem and of this sharpened completeness theorem: if a
proposition has a proof then, by the soundness theorem, it is valid in all
models, hence, by the sharpened completeness theorem, it has a cut
free proof.

Olivier Hermant has shown that such a model-based method
could be used to prove cut elimination for a large class of theories
in deduction modulo \cite{HermantThese,Hermant2005}.  His
sharpened completeness theorems are uniform: they proceed by constructing
a model ${\cal M}$ such that for all $A$, if $A$ is valid in
${\cal M}$, then $A$ has a cut free proof.  Notice that, when such a
uniform sharpened completeness theorem is used, only one instance of the
soundness theorem is needed in the proof of the cut elimination theorem:
that corresponding to the model ${\cal M}$ given by the sharpened
completeness theorem.  Provable propositions are valid in this
particular model, hence they have a cut free proof.

\section{Reductions}

The reduction-based methods prove cut elimination theorems by
exhibiting an algorithm that transforms a proof that is not cut free,
into another that is ``closer'' to a cut free proof. This
transformation process, called a {\em reduction} of the proof, can be
iterated and if it terminates, it yields a cut free proof. 

One of the first reduction-based cut elimination proofs is the proof
of termination of proof reduction in arithmetic using Tait's method
\cite{Tait67}. The idea is to prove, by induction over proof
structure, that all proofs terminate, but to let this induction go
through, it is necessary, as usual in proofs by induction, to
strengthen the induction hypothesis and to prove that all proofs
verify a property stronger than termination, called {\em
reducibility}. This notion of reducibility is parametrized by the
proposition the proof is a proof of. Thus, in this proof, a notion
of {\em being a reducible proof of $A$} is defined by induction
over the structure of the proposition $A$, and then, the fact that all
proofs of $A$ are reducible proofs of $A$ is proved by induction over
proof structure.

An equivalent formulation uses the set of reducible proofs of $A$
instead of the predicate ``being a reducible proof of $A$''.  The fact
that the predicate ``being a reducible proof of $A$'' is defined by
induction on the structure of $A$ then rephrases as the fact that the
set of reducible proofs of $A \Rightarrow B$ is defined from the set
of reducible proofs of $A$ and that of reducible proofs of $B$, by
applying to these sets an binary function $\tilde{\Rightarrow}$, and
similarly for the other connectors and quantifiers.

\section{From reductions to models}

Bridging the gap between reduction-based methods and model-based
ones has required several steps. 

\subsection{Reducibility candidates}

The first has been the introduction, by Jean-Yves Girard, of the
notion of {\em reducibility candidate} in his reduction-based proof
of cut elimination for simple type theory \cite{Girard,Girard72}.

In this proof, to define the set of reducible proofs of some
proposition $A$, it is necessary to quantify over the sets of reducible
proofs of all propositions $B$. Thus, a naive attempt leads to a
circular definition and a way to avoid this circularity is to
introduce {\em a priori} a set of sets of proofs, the set of {\em
reducibility candidates}, and to quantify over all such reducibility
candidates instead. Then, it is possible to define the set of
reducible proofs of all propositions and it happens {\em a posteriori}
that these sets are reducibility candidates.

Introducing this set of reducibility candidates, Jean-Yves Girard has
defined the place where sets of reducible proofs live.  Michel Parigot
\cite{Parigot} has shown later that this set of reducibility candidates
could be defined in a simple way, as the smallest set of sets of proofs
closed by the operations $\tilde{\Rightarrow}$, $\tilde{\forall}$, ...

\subsection{Reducibility candidates as truth values}

In the cut elimination proofs for various formulations of type theory,
in particular in Benjamin Werner's proof of cut elimination for the
{\em Calculus of inductive constructions} \cite{Werner}, the
similarity of this assignment of a set of proofs to each
proposition and the assignment of a truth value to each proposition in a
model was noticed.

In particular, it is interesting to remark that the model theoretic
notation $\llbracket A \rrbracket$ has progressively replaced the
notation $R_A$ for the set of reducible proofs of a proposition $A$
and that the expressions ``the interpretation of $A$'' and ``the
denotation of $A$'' have progressively replaced the expression ``the
set of reducible proofs of $A$''.

\subsection{Introducing domains}

In \cite{DowekWerner}, we have proposed, together with Benjamin Werner, a
reduction-based cut elimination proof for a large class of theories in
deduction modulo, characterized by the fact that they have a reducibility 
candidate valued model. 

For instance, to prove cut elimination modulo the rule $P
\longrightarrow (Q \Rightarrow R)$, we build a model where each
proposition is interpreted by a reducibility candidate and where the
rule $P \longrightarrow (Q \Rightarrow R)$, is valid, {\em i.e.} where
both sides of this rule have the same denotation.  We interpret first
the proposition symbols $Q$ and $R$ by any reducibility candidate, for
instance by the candidate $\tilde{\top}$, containing all the strongly
terminating proofs. Then, we interpret the proposition symbol $P$ by
the candidate $\llbracket Q \rrbracket~\tilde{\Rightarrow}~\llbracket
R \rrbracket$, in this example
$\tilde{\top}~\tilde{\Rightarrow}~\tilde{\top}$. Then, by
construction, we have $\llbracket P \rrbracket = \llbracket Q
\rrbracket~\tilde{\Rightarrow}~\llbracket R \rrbracket$, {\em i.e.} 
the rule $P \longrightarrow (Q \Rightarrow R)$ is valid in this
model. And the existence of such a model is sufficient to prove the
strong termination of proof reduction in this theory, as we can prove
that all proofs of a proposition $A$ are elements of the candidate
$\llbracket A \rrbracket$.

In deduction modulo, unlike in type theory, the terms of the theory
and the proof-terms are entities of different kinds, as well as the
sorts of the language and the propositions. Thus, it was natural to
interpret not only propositions, using reducibility candidates for
truth values, but also terms.  For instance, to build a model of the
rule $P(f(x)) \longrightarrow (P(x) \Rightarrow R)$, we can first
chose an term interpretation domain, for instance the set
${\mathbb N}$ of natural numbers, and an interpretation for the
function symbol $f$, for instance the function $n \mapsto n+1$. Then,
we interpret the proposition symbol $R$ by any reducibility candidate,
for instance by $\tilde{\top}$ and then the predicate symbol $P$ by
the function $\alpha$ mapping each natural number to a reducibility
candidate, defined by induction as follows: $\alpha(0)$ is any
candidate, for instance $\tilde{\top}$, and $\alpha(n+1) =
\alpha(n)~\tilde{\Rightarrow}~\llbracket R \rrbracket$. Then, it is
easy to check that for all valuations $\phi$, $\llbracket P(f(x))
\rrbracket_{\phi} = \llbracket P(x) \Rightarrow R \rrbracket_{\phi}$
and hence that the rule $P(f(x)) \longrightarrow (P(x) \Rightarrow R)$
is valid in this model.

Introducing this term interpretation domain simplified our cut
elimination proofs, in particular because, instead of defining the
interpretation of a predicate symbol as a function mapping terms to
truth values, it was possible to decompose this function in two steps
and first interpret the terms and then define the interpretation of a
predicate symbol as a function mapping elements of the term interpretation
domain to truth values, as it is usual in models.

This way, term interpretation domains were introduced
in reduction-based cut elimination proofs and this materialized in a
notion of reducibility candidate valued model, called {\em pre-models}.

\subsection{Truth values algebras and super-consistency}

As already said, the usual notion of bi-valued model can be extended
to notions where the truth values form a boolean algebra or a Heyting
algebra. This raises the question of the possibility to view
pre-models as such Heyting algebra valued models, {\em i.e.} the
question of the possibility to define an order on the set of
reducibility candidates that makes it a Heyting algebra.
Unfortunately, this is not possible, as in all Heyting algebras we
have $(\tilde{\top}~\tilde{\Rightarrow}~\tilde{\top}) = \tilde{\top}$,
but not in the algebra of reducibility candidates.  Thus, to include
the algebra of reducibility candidates, the notion of Heyting algebra
had to be generalized to a notion of {\em Truth values algebra}
\cite{TVA}.

What relations and operations should a set be equipped with, in order
to be used as a set of truth values? In fact, it does not need
to be equipped with an order relation. All that is needed is a family
of operations $\tilde{\Rightarrow}$, $\tilde{\forall}$,
$\tilde{\wedge}$, ... so that propositions can be interpreted and a
notion of {\em positive truth value} to characterize valid
propositions.  For the soundness theorem to hold, this set of positive
truth values must be closed by deduction rules. For instance, if
$a~\tilde{\Rightarrow}~b$ and $a$ are positive truth values, then $b$
also must be a positive truth value.

Thierry Coquand has suggested, in a personal communication, that, in
such an algebra, it is always possible to define a relation $\leq$ by
$a \leq b$ if $a~\tilde{\Rightarrow}~b$ is a positive truth value. And
he has noticed that a truth value algebra equipped with such a
relation verifies all the properties of Heyting algebras except one:
the antisymmetry of the relation $\leq$. Thus, truth values algebras
can alternatively be defined as pre-ordered structures with greatest
lower bounds, least upper bounds and relative complementation. Unlike
in ordered structures, greatest lower bounds and least upper bounds
are not unique in pre-ordered structures and, besides the pre-order,
the operations $\tilde{\Rightarrow}$, $\tilde{\forall}$,
$\tilde{\wedge}$, ... must be given in the definition of the algebra,
as it has to be specified which greatest lower bound of $a$ and $b$
the element $a~\tilde{\wedge}~b$ is.

It is well-known that the relation defined on propositions by $A \leq B$
if $A \Rightarrow B$ is provable is reflexive and transitive, but that
to make it antisymmetric and define the Lindenbaum algebra of a
language, it is necessary to quotient the set of
propositions by the relation $\simeq$ defined by $A \simeq B$ if $A
\Leftrightarrow B$ is provable. An alternative ``solution'' is to drop
this antisymmetry requirement.

The set of reducibility candidates equipped with the operations
$\tilde{\Rightarrow}$, $\tilde{\forall}$, $\tilde{\wedge}$, ...  is a
truth values algebra and the models valued in this algebra are exactly
the pre-models we had defined with Benjamin Werner.  

Surprisingly, in these constructions, no specific properties of the
algebra of reducibility candidates were used. Thus, they generalize to
all truth values algebras. This has lead to introduce a notion of {\em
super-consistency}: a theory is super-consistent it if has a ${\cal
B}$-valued model, not only for one, but for all truth values algebras
${\cal B}$.  For instance, in any truth value algebra ${\cal B}$, we
can build a model of the rule $P \longrightarrow (Q \Rightarrow R)$ by
interpreting $Q$ and $R$ by the truth value $\tilde{\top}$ and $P$ by
the truth value $\tilde{\top}~\tilde{\Rightarrow}~\tilde{\top}$. Thus,
the theory formed with the rewrite rule $P \longrightarrow (Q
\Rightarrow R)$ is super-consistent.  In the same way, we can prove
that arithmetic and simple type theory are super-consistent.

Super-consistency is a model theoretic sufficient condition for
termination of proof reduction. Whether this condition is necessary is
still an open problem.

\subsection{From super-consistency to model-based cut elimination proofs}

We have noticed with Olivier Hermant that the proof that
super-consistency implies the termination of proof reduction can be
simplified, if we restrict the goal to prove that super-consistency
implies that all provable propositions have a cut free proof
\cite{DowekHermant}.  In this 
case, we do not need to establish a property of proofs but merely a
property of propositions and sequents. Thus, instead of using an algebra whose
elements are sets of proofs, we can use a simpler algebra
whose elements are sets of sequents: in a
reducibility candidate, each proof collapses to its conclusion.

Tait's lemma, that if $\pi$ is a proof of a proposition $A$, then it
is an element of the set $\llbracket A \rrbracket$ of reducible proofs
of $A$, collapses to the fact that if a sequent $\Gamma \vdash A$ is
provable then it is a element of $\llbracket A \rrbracket$.  This
model can further be transformed into a model where truth values are
sets of contexts, that form a Heyting algebra, in such a way that if
$\Gamma \in \llbracket A \rrbracket$ then $\Gamma \vdash A$ has a cut
free proof, and Tait's lemma rephrases as the fact that if a sequent
$\Gamma \vdash A$ is provable then $\Gamma \in \llbracket A
\rrbracket$.

It is possible to prove that if a sequent $\Gamma \vdash A$ is valid
in this model, then $\Gamma \in \llbracket A \rrbracket$ and Tait's
lemma boils down to the fact that if a sequent $\Gamma \vdash A$ is
provable then it is valid in this model. Like in model-based proofs,
this is just the instance of the soundness lemma corresponding to this
model. And indeed the fact that if a sequent $\Gamma \vdash A$ is
valid in this model, then $\Gamma \in \llbracket A \rrbracket$ and
thus $\Gamma \vdash A$ has a cut free proof is a uniform sharpened
completeness theorem.

\section*{Conclusion}

Step by step, the gap between reduction-based methods and model-based
methods to prove cut elimination has been bridged and, along the way,
cut eliminations has been proved for new theories. Both types of
proofs can be decomposed in two steps, first a proof that
the theory of interest is super-consistent, then a proof, independent
of the theory of interest, that
super-consistency implies cut elimination. The only difference is
in the choice of the truth value algebra used to deduce 
cut elimination from super-consistency: the algebra of reducibility
candidates to prove the termination of proof reduction, the simpler
algebra of sequents to construct a model where validity
implies cut free provability.

Once this convergence is achieved, 
several directions may be worth exploring. First, we may
want to deduce directly proof theoretical results from
super-consistency, without proving cut elimination first. Second, we
may want to extend the notion of super-consistency to type theory. The
recent \cite{CousineauDowek} that relates deduction modulo and type
theory may be a good starting point.


\begin{thebibliography}{99.}

\bibitem{Allali}
L.~Allali.
Algorithmic equality in Heyting arithmetic modulo.
{\em Higher Order Rewriting}, 2007.

\bibitem{Andrews} 
P.B.~Andrews.  
Resolution in type theory.  
{\em The Journal of Symbolic Logic}, 36(3):414--432, 1971.

\bibitem{KirchnerBraunerHoutmann}
P.~Brauner, C.~Houtmann, and C.~Kirchner.
Superdeduction at work. 
{\em Rewriting, Computation and Proof, Essays dedicated to Jean-Pierre
  Jouannaud on the occasion of his 60th birthday}, 
Lectures Notes in Computer Science 4600, Springer,
132--166, 2007.

\bibitem{BraunerDowekWack}
P.~Brauner, G.~Dowek, and B.~Wack.
Normalization in supernatural deduction and in deduction modulo.
Manuscript, 2007.

\bibitem{CousineauDowek} 
D.~Cousineau and G.~Dowek.
Embedding Pure Types Systems in the lambda Pi-calculus modulo,
{\em Typed Lambda calculi and Applications},
Lecture Notes in Computer Science 4583, Springer, 
102--117, 2007.    

\bibitem{Crabbe74}
M.~Crabb\'e.
Non-normalisation de la th\'eorie de Zermelo.
Manuscript, 1974.

\bibitem{Crabbe91}
M.~Crabb\'e.
Stratification and cut-elimination.
{\em The Journal of Symbolic Logic}, 56(1):  213--226, 1991.

\bibitem{DeMarcoLipton} 
M.~De~Marco and J.~Lipton.  
Completeness and cut-elimination in the
intuitionistic theory of types.  
{\em Journal of Logic and Computation}, 15:821--854, 2005.

\bibitem{foldunfold}
G.~Dowek.
About folding-unfolding cuts and cuts modulo.
{\em Journal of Logic and Computation}, 11(3):419-429, 2001.

\bibitem{TVA}
G.~Dowek, 
Truth values algebras and proof normalization, {\em TYPES 2006}, 
Lectures Notes in Computer Science 4502, Springer, 2007.

\bibitem{DHK} 
G.~Dowek, T.~Hardin, and C.~Kirchner.  
Theorem proving modulo.  
{\em Journal of Automated Reasoning}, 31:32--72, 2003.

\bibitem{DHKHOL} 
G.~Dowek, T.~Hardin, and C.~Kirchner.  
HOL-lambda-sigma: an intentional first-order expression of
higher-order logic.
{\em Mathematical Structures in Computer Science}, 11:1--25, 2001.

\bibitem{DowekHermant}
G.~Dowek and O.~Hermant.  
A simple proof that super-consistency implies cut elimination.
{\em Rewriting techniques and applications},
Lecture Notes in Computer Science, 4533, 
96--106, 2007.

\bibitem{DowekWerner}
G.~Dowek and B.~Werner.  Proof normalization modulo.  {\em The Journal
of Symbolic Logic}, 68(4):1289--1316, 2003.

\bibitem{Peano}
G.~Dowek and B.~Werner. Arithmetic as a theory modulo. 
{\em Term rewriting and applications}, 
Lecture Notes in Computer Science 3467, Springer, 423--437, 2005.

\bibitem{Ekman}
J.~Ekman.
{\em Normal proofs in set theory}. Doctoral thesis, 
Chalmers university of technology and University of G\"{o}teborg,
  1994.

\bibitem{Girard}
J.-Y. Girard.  Une extension de l'interpr\'etation de {G}\"odel \`a
l'analyse, et son application \`a l'\'elimination des coupures dans
l'analyse et la th\'eorie des types.  {\em 2$^{\mbox{nd}}$
  Scandinavian Logic Symposium}, Noth Holland, 63--92, 1971.

\bibitem{Girard72}
J.-Y. Girard, 
{\em Interpr\'etation fonctionnelle et \'elimination des coupures dans 
l'arithm\'etique d'ordre sup\'erieur.}
Doctoral thesis, 
Universit\'e de Paris 7, 1972.

\bibitem{Hallnas}
L.~Halln\"{a}s.
{\em On normalization of proofs in set theory}.
Doctoral thesis, University of Stockholm, 1983.

\bibitem{HermantThese}
O.~Hermant.  {\em M\'ethodes s\'emantiques en d\'eduction modulo.}
Doctoral Thesis, Universit\'e de Paris 7, 2005.

\bibitem{Hermant2005}
O.~Hermant.  Semantic cut elimination in the intuitionistic sequent
calculus.  {\em Typed Lambda Calculi and Applications}, 
Lectures Notes in Computer Science 3461, Springer,
221--233, 2005.

\bibitem{Okada}
M.~Okada.  A uniform semantic proof for cut elimination and
completeness of various first and higher order logics.  {\em
Theoretical Computer Science}, 281:471--498, 2002.

\bibitem{Parigot}
M.~Parigot.
Strong normalization for the second order classical natural
deduction.
{\em Logic in Computer Science}, 39--46, 1993.

\bibitem{Prawitz} 
D.~Prawitz. Hauptsatz for higher order logic. {\em The Journal of
Symbolic Logic}, 33:452--457, 1968.

\bibitem{PrawitzND}
D.~Prawitz.
{\em Natural Deduction. A Proof-Theoretical Study.}
Almqvist and Wiksell, 1965.

\bibitem{Tait66}
W.~W. Tait.  A non constructive proof of {G}entzen's {H}auptsatz for
second order predicate logic.  {\em Bulletin of the American
Mathematical Society}, 72:980--983, 1966.

\bibitem{Tait67} W.~W. Tait. Intentional interpretations of
functionals of finite type {I}.  {\em The Journal of Symbolic Logic},
32:198--212, 1967.

\bibitem{Takahashi}
M.~o. Takahashi. A proof of cut-elimination theorem in simple type theory.
{\em Journal of the Mathematical Society of Japan}, 19:399--410, 1967.

\bibitem{Wack}
B.~Wack.
{\em Typage et d\'eduction dans le calcul de r\'e\'ecriture.}
Doctoral Thesis, Universit\'e Henri Poincar\'e Nancy 1, 
2005.

\bibitem{Werner}
B.~Werner.
{\em Une th\'eorie des constructions inductives.}
Doctoral Thesis, Universit\'e de Paris 7, 1994.

\end{thebibliography}
\end{document}